\documentclass[aps,prd,twocolumn,floatfix,nofootinbib,showpacs]{revtex4}
\usepackage{color}
\usepackage{amsmath,amssymb}
\usepackage{graphicx}
\usepackage{latexsym}

\newcommand{\eq}[1]{(\ref{#1})}
\newcommand{\nn}{\nonumber}

\newcommand{\del}{\partial}

\newcommand{\asymeq}{\underset{asym}{\simeq}}

\DeclareMathOperator{\diag}{diag}

\renewcommand{\thefootnote}{\fnsymbol{footnote}}

\begin{document}

\title{Analytic Study for the String Theory Landscapes via Matrix Models}

\author{
Chuan-Tsung Chan$^{a,b}$, Hirotaka Irie$^{b,c}$ and Chi-Hsien Yeh$^{b,d}$}
\affiliation{
$^{a}$Department of Physics, Tunghai University, Taichung 40704, Taiwan\\
$^{b}$National Center for Theoretical Sciences, 
National Tsing-Hua University, Hsinchu 30013, Taiwan \\
$^{c}$Yukawa Institute for Theoretical Physics, 
Kyoto University, Kyoto 606-8502, Japan \\
$^{d}$Department of Physics and Center for Theoretical Sciences, 
National Taiwan University, Taipei 10617, Taiwan 
}
\renewcommand{\thefootnote}{\arabic{footnote}}
\date{June 2012; Revised: November 2012}

\preprint{YITP-12-49}

\begin{abstract}
We demonstrate a first-principle analysis of the string theory landscapes 
in the framework of non-critical string/matrix models. 
In particular, we discuss non-perturbative instability, decay rate and the true vacuum of perturbative string theories. As a simple example, we argue that the perturbative string vacuum of pure gravity {\em is} stable; 
but that of Yang-Lee edge singularity is {\em inescapably} a false vacuum. 
Surprisingly, most of perturbative minimal string vacua are unstable, and their true vacuum mostly does not suffer from non-perturbative ambiguity. Importantly, we observe that the instability of these tachyon-less closed string theories is caused by {\em ghost D-instantons} (or ghost ZZ-branes), the existence of which is determined only by non-perturbative completion of string theory. 
\end{abstract}
\maketitle

\section{Analytic aspects of the string theory landscape}

The string theory landscape is a space of vacua in string theory, 
which hopefully includes the standard model in four dimension. 
Despite of its importance, its progress is mainly in its statistical aspects \cite{Landscape}; and 
little is known about analytic structures of the space. By ``analytic structures'' we mean general interrelationship among distinct perturbative string-theory vacua. Therefore, appearance of vacua in the landscape, relative stability/decay rate of vacua and identification of the true vacuum are included. In this short letter, by use of the non-perturbative completion, we demonstrate a prescription to extract analytic structures of the string theory landscapes from perturbative string theory. 

The free energy of perturbative string theory, $\mathcal F(g)$, is an asymptotic series and is calculated from world-sheet conformal field theory \cite{StringText}: 
\begin{align}
\mathcal F(g) \simeq \sum_{n=0}^\infty g^{2n-2} \mathcal F_n + \sum_{I} \theta_I g^{\gamma_I}&\exp\Bigl[\frac{1}{g} \sum_{n=0}^\infty g^{n} \mathcal F^{(I)}_n\Bigr]  + \mathcal O(\theta^2). \label{AsymExpWithInstantons}
\end{align}
The nonperturbative corrections are usually provided by D-instantons, 
i.e.~their leading contributions, $\mathcal F^{(I)}_0$, are identified as D-instanton action $\mathcal S_I = - \frac{1}{g}\mathcal F^{(I)}_0$ \cite{ShenkerPolchinski,Reloaded}. The over-all coefficient $\theta_I$ for each instanton is called D-instanton fugacity \cite{David,fy,HHIKKMT},
which has no corresponding worldsheet observable. 
Usually, we assume that the D-instanton action is positive: $\mathcal S_I>0$. However, a negative-action partner of the instanton, $\mathcal S_{I_{\rm gh}} = - \mathcal S_I<0$ has also been observed \cite{SeSh} in non-critical string theory \cite{NonCriticalStrings}. They are then generally defined as ghost D-branes (or ghost D-instantons) in (non-)critical string theory \cite{GhostDbranes}. 
However, such a D-brane was not seriously taken into account, 
since it contradicts with perturbation theory. 
Existence of the D-branes is discussed very recently 
mainly in resurgent analysis \cite{Resurgent,KMR,ASV} and it was found that 
these branes must be generally encoded in non-perturbatively completion of string theory.%
\footnote{It is shown that ``multi instanton-ghost-instanton sectors'' have discrepancy with worldsheet predictions in the sense of $\mathcal F^{(n|m)} \neq \mathcal F^{(n-m|0)}$ \cite{KMR,ASV}, and this is a main objection to identifying it as ``ghost D-branes''. 
However we insist on using the terminology because, according to the free-fermion analysis \cite{fy,fisfim}, multi ghost-instanton sectors $\mathcal F^{(0|m)}$ are simplify obtained by flipping the sign of the ZZ-brane boundary state operators in the multi-instanton sectors $\mathcal F^{(m|0)}$ in all-order perturbation theory. }
In this letter, we shall see how these ghost D-instantons play 
a role in formulating ``analytic structures'' of the string theory landscape. 

Since the actions of ghost D-instantons are negative (their masses are negative), 
they are no longer ``corrections'' to perturbation theory; they are rather {\em indication of non-perturbative instability of the perturbative vacuum} \cite{KMR}. 
However, this is a cause of confusions, because ``in principle, the ghost partner is defined for every D-brane, but it does not necessarily mean that the string theory is unstable.'' 
In fact, it is non-trivial to know {\em which ghost D-instantons are allowed (or not allowed) to appear in the spectrum.} Naively, this information is given by physics of the D-instanton fugacity $\{\theta_I\}_I$. However, it is subtle to directly deal with $\{\theta_I\}_I$ since they are coefficients of exponentially small corrections which are supposed to be negligible in asymptotic expansions (See e.g.~\cite{MarinoLecture}). Therefore, we should first grasp complete information of D-instanton fugacity. In the following, we shall see that, once one can control the information of D-instanton fugacity, one can quantitatively extract most of analytic aspects of the string theory landscapes, including {\em metastability, its decay rates} and {\em the true vacuum}.

{\em The completion and fugacity}\quad 
It is known that perturbative amplitudes, including instanton corrections, in various solvable string theories are obtained by the information of spectral curves, especially with topological recursions \cite{EynardOrantin,EynardMarino}. In particular, all-order asymptotic expansion of Eq.~\eq{AsymExpWithInstantons} is explicitly shown in \cite{EynardMarino}, with fugacity remaining free parameters. Then, for completion of the non-perturbative information in the asymptotic expansion, there are mainly studied two ways to control fugacity: one is resurgent analysis (e.g.~\cite{Resurgent,KMR,ASV,MarinoLecture}); and the other is isomonodromy analysis (for mathematical developments on isomonodromy theory \cite{RHcite,RHPIIcite,ItsBook}; and with matrix models \cite{Moore,FIK,CIY2,CIY3}). The former is based on connection formula (or Stokes phenomena) for analytic continuation of $g$; and the latter is based on {\em Stokes phenomena of the Baker-Akhiezer (shortly BA) functions on the spectral curves}. Here we explore analytic aspects of the landscape from the latter approach. 

\section{Riemann-Hilbert problem for the Baker-Akhiezer functions}

\begin{figure}[htbp]
\begin{center}
\includegraphics[scale=0.6]{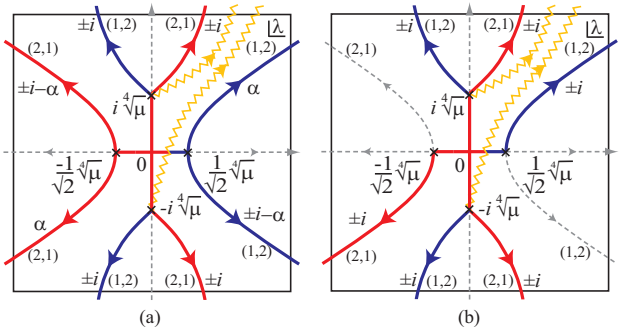}
\end{center}
\caption{\footnotesize The Deift-Zhou network for pure-gravity. (a) General solutions to the non-perturbative completion with two-cut boundary condition. (b) A solution with the single-line condition. There is also another solution obtained by the reflection with respect to real axes. }
\label{DZgraph23}
\end{figure}

For a given spectral curve $F(P,Q)=0$ with a symplectic coordinate $(P,Q)$, 
we define the BA function as follows: 

1) We define a function $\varphi(\zeta)$ (called string-background) as 
\begin{align}
\varphi(\zeta) = \underset{1\leq j\leq k}{\diag} \bigl(\varphi^{(j)}(\zeta)\bigr),\quad 
\varphi^{(j)}(\zeta) = \int^\zeta dP\, Q^{(j)}(P), \label{stringbackground}
\end{align}
where $\{Q^{(j)}(P)\}_{j=1}^k$ are branches of the algebraic equation, the number of which is an integer, $k$. 
The function $\varphi(\zeta)$ is a rational function on the curve and, without loss of generality,  it may have poles at $\zeta = \infty$ and $\zeta = \zeta_a$ $(a=1,2,\cdots,M-1)$ in the following sense:
\begin{align}
\varphi(\zeta) &\sim \sum_{n=1}^{r_0} \varphi_{-n} \lambda^{n}+ \mathcal O(\frac{1}{\lambda}),\quad \zeta = \lambda^{\hat p_0}\to \infty, \nn\\
\varphi(\zeta) &\sim \sum_{n=1}^{r_a}\frac{\varphi_{-n}(\zeta_a)}{\lambda^{n} }+ \mathcal O(\lambda),\quad \zeta = \zeta_a + \lambda^{\hat p_a}\to \zeta_a,
\label{ExpOfStringBackground}
\end{align}
with $a=1,2,\cdots,M-1$. Here $\{\hat p_a\}_{a=0}^{M-1}$ are proper integers and $\{r_a\}_{a=0}^{M-1}$ are the Poincar\'e indices. 
The BA function $\Psi(\zeta)$ is then a $k\times k$ matrix-valued sectional holomorphic function of $\zeta \in \mathbb C^*\backslash \mathcal K$ as 
\begin{align}
\Psi(\zeta) = Z(\zeta)e^{\varphi(\zeta)} 
\prod_{a=0}^{M-1}(\zeta-\zeta_a)^{\nu_a/\hat p_a} \equiv Z(\zeta)e^{\bar \varphi(\zeta)},
\end{align}
where $\mathcal K$ is a collection of connected line elements, $\mathcal K = \bigcup_m \mathcal K_m$ equipped with a direction (e.g.~Fig.~\ref{DZgraph23} and Fig.~\ref{DZgraph25}), and $\zeta_0 = 0$. We often put $\hat p_0 = \hat p$, $r_0 = r$ and $\nu_0 = -\nu$. 

Note that the line elements of the graph $\mathcal K$ flow from the poles of $\varphi(\zeta)$ (i.e.~essential singularities of the BA function) and are drawn along anti-Stokes lines, ${\rm Re}\bigl[(\varphi^{(j)}(\zeta)-\varphi^{(l)}(\zeta))e^{-i\theta}\bigr] = 0$, with a proper $\theta$ so that the graph $\mathcal K$ attaches to saddle points, $\partial_\zeta\bigl(\varphi^{(j)}(\zeta)-\varphi^{(l)}(\zeta)\bigr)=0$. 

2) For each segment of the graph, $\mathcal K_m$, a $k\times k$ matrix $S_m$ is assigned (which is called a {\em Stokes matrix}) and then the BA function has discontinuity along the segment, 
\begin{align}
\Psi(\zeta+\epsilon)=\Psi(\zeta-\epsilon) S_m,\qquad \zeta\in \mathcal K_m, 
\label{StokesJump}
\end{align}
where $\epsilon$ directs to the left-hand side of the segment $\mathcal K_m$. 
In particular, at the poles of $\varphi(\zeta)$, there are a number of lines (as in Eq.~\eq{MonodromyFree}) and the Stokes matrices are defined so that the BA function has standard asymptotic expansion around them: 
\begin{align}
\Psi(\zeta) &\asymeq \Bigl[I_k + \sum_{n=1}^\infty \frac{Z_n}{\lambda^n} \Bigr] e^{\bar \varphi(\zeta)}\quad (\zeta= \lambda^{\hat p}\to \infty), \nn\\
\Psi(\zeta) &\asymeq \Bigl[I_k + \sum_{n=1}^\infty \lambda^n Z_n(\zeta_a) \Bigr] e^{\bar \varphi(\zeta)} E_a\quad \nn\\
& \qquad\qquad\qquad\qquad\quad\,\,  (\zeta = \zeta_a + \lambda^{\hat p_a}\to \zeta_a),
\label{AsymExpBAatSingularity}
\end{align}
with $\det E_a \neq 0$ and $a=1,2,\cdots,M-1$. Note that the expansion \eq{AsymExpBAatSingularity} does not depend on the direction of $\zeta\to \zeta_0$, and therefore this requires the standard form for the Stokes matrices \eq{StokesJump}. 

Note that if one flips the direction of a line $\mathcal K_m$ then the matrix is replaced by its inverse, $S_m^{-1}$. At a junction of lines, they satisfy a conservation equation: 
\begin{align}
\includegraphics[scale = 0.7]{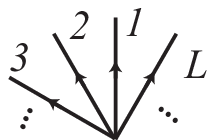} \qquad \Leftrightarrow \qquad S_1S_2\cdots S_L = I_k. \label{MonodromyFree}
\end{align}
These are the basic algebra of the Stokes matrices. Obtaining explicit solutions to the algebra is a first non-trivial preparation for the Riemann-Hilbert (shortly RH) calculus, and some recent progress for general $k$ and $r$ can be found in \cite{CIY2,CIY3}.

Importantly, the Stokes matrices $S_m$ in Eq.~\eq{StokesJump} are independent from $\zeta$ of $\mathcal K_m$, which means that the graph $\mathcal K$ is topological and one can deform it continuously. In addition, we assume that the Stokes matrices are independent from the deformations of the leading Laurent coefficients in Eq.~\eq{ExpOfStringBackground}, i.e.~$\{\varphi_{-n}\}_{n=1}^r$ and $\{\varphi_{-n}(\zeta_a)\}_{n=1}^{r_a}{}_{a=1}^{M-1}$. They are the isomonodromy deformations 
which guarantees the integrable hierarchy (e.g.~KP/Toda hierarchy according to the spectral curves \cite{KPToda}) and string equations of the system, which means that perturbative results coincidence with that of the topological recursions. The partition function of matrix models is then given by the $\tau$-function of the integrable hierarchy 
(See e.g.~\cite{FKN2}).

3) The sectional holomorphic function $Z(\zeta)$ is uniquely fixed by giving the Stokes matrices of jump relations \eq{StokesJump}. In fact, $Z(\zeta)$ is calculable by solving the Riemann-Hilbert integral equation (See e.g.~\cite{ItsBook}) around $\zeta = \lambda^{\hat p} \to \infty$: 
\begin{align}
\widetilde Z(\lambda) = I_k + \int_{\mathcal K} \frac{d \xi}{2\pi i} \frac{\widetilde Z(\xi-\epsilon)(G(\xi)-I_k)}{\xi-\lambda},\label{RHintegral}
\end{align}
along $\mathcal K$. Here $\widetilde Z(\lambda) \equiv Z(\lambda^{\hat p})$ and $G(\lambda)$ is a sectional holomorphic function along $\mathcal K$, defined by $G(\lambda) = e^{\bar \varphi(\lambda^{\hat p})} S_m e^{-\bar \varphi(\lambda^{\hat p})}$ ($\lambda \in \mathcal K_m$; $m=0,1,\cdots$).

 We should note that by this procedure one observes that fugacity is given by Stokes multipliers of $\{S_m\}_m$ mostly related linearly (See e.g.~\cite{ItsBook}). Importantly this allows us to obtain the connection rules for analytic continuation of integrable flows (including string coupling $g$). In this sense, the information of graph and matrices, $\hat {\mathcal K} \equiv \bigcup_m(\mathcal K_m,{S_m})$ has all the information of D-instanton fugacity of Eq.~\eq{AsymExpWithInstantons}. We shall refer to $\hat {\mathcal K}$ as {\em the Deift-Zhou (or DZ) network} \cite{DeiftZhou} (by following the recent naming fashion \cite{SpectralNetworks}). This is how we control the fugacity. 

4) In the RH approach, if one fixes the integrable flows $\{\varphi_{-n}(\zeta_a)\}_{n=1}^{r_a}{}_{a=1}^M$ and Stokes matrices $\{S_m\}_m$, every information is determined. In particular, the BA function $\Psi(\zeta)$ is not changed by any deformations of the spectral curve, $F(P,Q)=0 \to \widetilde F(P,Q)=0$ (i.e.~$\varphi(\zeta)\to\, \widetilde \varphi(\zeta)$), such that the resulting string-background $\widetilde \varphi(\zeta)$ of Eq.~\eq{stringbackground} does not change the singular structure \eq{ExpOfStringBackground}. In other words, this is just a matter of how divide the BA function $\Psi(\zeta)$ into $Z(\zeta)$ and $\bar \varphi(\zeta)$. In this sense, the RH approach can be interpreted as an (off-shell) background independent formulation of string theory \cite{CIY2}. 

By this fact, we define the string theory landscape $\mathfrak L_{\rm str}$ by the moduli space of spectral curves $F(P,Q)=0$ which preserves the pole structure (i.e.~integrable flows) of $\varphi(\zeta)$ in Eq.~\eq{ExpOfStringBackground}. Schematically, we define it as a set of string-background $\varphi(\zeta)$: 
\begin{align}
\mathfrak L_{\rm str} = \bigl\{\varphi(\zeta);  \text{keeping Eq.~\eq{ExpOfStringBackground}}\bigr\},
\end{align}
and the potential of the landscape is determined by the RH integral equation \eq{RHintegral}. 

Note that the background independence of non-critical string theory was first explicitly shown in the topological recursions \cite{EynardMarino} i.e.~within perturbation theory; however for the potential picture of landscape we need to know the non-perturbative completion of the string theory. In the following, as a first non-trivial example for the analytic aspects of the landscape, we shall show how meta-stability, decay rates and true vacuum are obtained with the information of fugacity/network which are controlled as above. 

\section{Cases of minimal string theory}
We now consider minimal string theory \cite{NonCriticalStrings}, as an example described by matrix models/spectral curves \cite{DSL}. The spectral curve of $(p,q)$ minimal string theory is given by 
\begin{align}
F(\zeta,Q) = T_p(Q/\beta \mu^{q/2p})-T_q(\zeta/\sqrt{\mu})=0,
\label{SpectralCurveMinimalString}
\end{align}
with Chebyshev polynomials of the first kind $T_n(\cos \theta) =\cos n\theta$ \cite{Kostov1,SeSh}. Therefore, $\varphi(\zeta) (\equiv \varphi_{\rm mstr}(\zeta))$ in this background is given as $\varphi_{\rm mstr}^{(j)}(\zeta) = \varphi_{\rm mstr}^{(1)}(e^{-2\pi i \frac{j-1}{p}}\zeta)$ with $\varphi_{\rm mstr}^{(1)}(\zeta) = \beta \mu^{\frac{q+2}{4}}\int^{\zeta/\sqrt{\mu}} dx \, T_{q/p}(x)$. 
That is, $(p,q)$ minimal string theory is $p\times p$ isomonodromy systems (i.e.~$k=p$) with only one essential singularity at $\zeta = \infty$
of Poincar\'e index $r=p+q$ (we put $\zeta = \lambda^{p}$ in Eq.~\eq{ExpOfStringBackground}). We put the monodromy $\nu_0$ as $\nu_0\, (=-\nu)=-\frac{p-1}{2}$, and this background also preserves $\mathbb Z_p$-symmetry in the sense of \cite{CIY2}. Here following the discussion \cite{CIY2}, we use the same notation. 

Around the singularity $\zeta\to \infty$, there are $2rp$ Stokes matrices $\{S_n\}_{n=0}^{2rp-1}$ of $p\times p$, and their algebraic relations \cite{ItsBook,CIY2} are expressed as 
\begin{itemize}
\item $\mathbb Z_p$-symmetry condition: \\ $S_{n+2r} = \Gamma^{-1} S_n \Gamma$ $\bigl(n=0,1,\cdots, 2rp-1\bigr)$,
\item Monodromy condition: \\ $S_0 S_1\cdots S_{2rp-1} = e^{\pi i (p-1)} I_p$, 
\item Hermiticity condition: \\ $S_n^*= \Delta\Gamma\, S_{(2r-1)p-n}^{-1}\, \Gamma^{-1}\Delta$ $\bigl(n=0,1,\cdots, 2rp-1\bigr)$. 
\end{itemize}
with $\Gamma = (\Gamma_{ij})_{1\leq i,j\leq p} = 
(\delta_{j,i+1}+\delta_{i,p}\delta_{j,1})_{1\leq i,j\leq p}$ and $\Delta = (\Delta_{ij})_{1\leq i,j\leq p} = (\delta_{i+j,p+1})_{1\leq i,j\leq p}$. 
For components of the matrices $\{S_n\}_{n=0}^{2pr-1}$, one should consult \cite{CIY2}. 

In addition, we consider the cases related to matrix models. The corresponding conditions for the Stokes matrices are known as the multi-cut boundary condition \cite{CIY2}. In particular, in the case of $(p,q)$ minimal string theory, the constraint is the same as $p$-cut critical points of the multi-cut matrix models \cite{MultiCut,FSSTMultiCut,CISY1,CIY1}.%
\footnote{
The constraint requires ``$p$ cuts'' (not one cut) around $\lambda \to \infty$ in the resolvent function of this $p\times p$ isomonodromy systems. It is because the spectral parameter $\lambda (= \zeta^{1/p})$ creates $p$ copies of the physical cuts in $\mathbb Z_p$-symmetric way.}
A major difference from the previous cases \cite{CIY2,CIY3} is, however, that the Poincar\'e index is greater than the number of cuts: 
\begin{align}
r \,(= p+q) > k\, (=p), 
\end{align}
which greatly simplify the quantum integrable structure of the condition \cite{CIY3}. Therefore, for simplicity, we below consider the cases of $p=2$, i.e.~one-matrix models. Then we can completely solve the conditions and the Stokes matrices are given with $(L=1,2,\cdots)$ as
\begin{itemize}
\item[1)] \underline{$r=q+2=4L+1$ cases} $(m=1,2,\cdots, L)$
\begin{align}
S_{4m-5} = &
\begin{pmatrix}
1 & 0 \cr
0 & 1
\end{pmatrix},\quad 
S_{4m-3} = 
\begin{pmatrix}
1 & \alpha_{m} \cr
0 & 1
\end{pmatrix} \nn\\
S_{4L-1} = &
\begin{pmatrix}
1 & 0 \cr
\pm i & 1
\end{pmatrix},\quad 
S_{4L+1} = 
\begin{pmatrix}
1 & \pm i \cr
0 & 1
\end{pmatrix} \nn\\
S_{4(2L-m)+3} = &
\begin{pmatrix}
1 & 0 \cr
\alpha_{-m} & 1
\end{pmatrix},\quad 
S_{4(2L-m)+5} = 
\begin{pmatrix}
1 & 0 \cr
0 & 1
\end{pmatrix} \nn
\end{align}
with $\sum_{m=1}^L (\alpha_m+\alpha_{-m})= \pm i$. 
\item[2)] 
\underline{$r=q+2=4L+3$ cases} $(m=0,1,\cdots, L)$
\begin{align}
S_{4m-3} = &
\begin{pmatrix}
1 & 0 \cr
0 & 1
\end{pmatrix},\quad 
S_{4m-1} = 
\begin{pmatrix}
1 & 0\cr
\alpha_{m} & 1
\end{pmatrix}  \nn\\
S_{4L+1} = &
\begin{pmatrix}
1 & \pm i  \cr
0 & 1
\end{pmatrix},\quad 
S_{4L+3} = 
\begin{pmatrix}
1 & 0 \cr
\pm i  & 1
\end{pmatrix} \nn\\
S_{4(2L-m)+5} = &
\begin{pmatrix}
1 & \alpha_{-m} \cr
0 & 1
\end{pmatrix},\quad 
S_{4(2L-m)+7} = 
\begin{pmatrix}
1 & 0 \cr
0 & 1
\end{pmatrix} \nn
\end{align}
with $\alpha_0+\sum_{m=1}^L (\alpha_m+\alpha_{-m})= \pm i$. 
\end{itemize}
Note that $S_{2m}=I_2$ ($m\in \mathbb Z$) and the remaining Stokes matrices are 
obtained by the $\mathbb Z_p$-symmetric condition. 
In addition, the hermiticity condition is given as $\alpha_m = -\alpha_{-m}^*$. 
In the following, we show the Deift-Zhou networks and results of the RH calculus in the $p=2$ cases. 


\subsection{Meta-stability and decay rates of string theory}
\paragraph{$(p,q)=(2,3)$ minimal strings: Pure-gravity}\quad
Non-perturbative completion in the pure-gravity case 
are studied from various points of view \cite{David,FIK,Resurgent}. 
The DZ network of the spectral curve is drawn in Fig.~\ref{DZgraph23}-a. 
Note that the hermiticity condition, $\alpha-\alpha^* = \pm i$, requires that 
the free-energy is a real function. On the other hand, 
it is known that non-perturbative solutions of the matrix models 
should break hermiticity \cite{David}. This situation is recovered 
if one discards the hermiticity condition and insists that 
{\em contours in the DZ network attached to saddle points (i.e.~D-instantons) should be 
a single line with a uniform Stokes multiplier} 
(See Fig.~\ref{DZgraph23}-b). 
We here refer to this condition as {\em the single-line condition}, which corresponds 
to the standard solutions from matrix models. In fact, the result is following $(g\to 0,\, \mu >0)$: 
\begin{align}
\mathcal F \asymeq \Bigl[-\frac{4}{15} \frac{\mu^{\frac{5}{2}}}{g^2}+ \mathcal O(g^0)\Bigr] +
\frac{\mp i}{2} \Bigl[\frac{\sqrt{g}\,(1+\mathcal O(g))}{8 \sqrt{3^{\frac{3}{2}} \pi}\, \mu^{\frac{5}{8}}}\Bigr] e^{-\frac{8\sqrt{3}}{5 g}\mu^{\frac{5}{4}}},  
\end{align}
and gives {\em one half of} the fugacity obtained in \cite{HHIKKMT}. 
This value is natural by the same reason as \cite{Coleman} and has been also argued in \cite{Penner}. 
Here $u=g^2\mathcal F''(t)$, $g^2 u''+6(u^2+t)=0$ and $\mu=-t$. 

This result suggests an important implications: (Anti-Stokes) lines of the DZ networks correspond to ``the mean field path-integral of the many-eigenvalue system (discussed in \cite{David})''. Therefore, the DZ networks are a remnant of {\em path-integral in string theory}. If this is so, with the standard prescription in QM/QFT systems \cite{Coleman}, one expects that {\em meta-stability} and {\em decay rates} can be discussed in string theory. For further discussions, we consider minimal string of the Yang-Lee edge, $(p,q)=(2,5)$. 


\begin{figure}[htbp]
\begin{center}
\includegraphics[scale=0.6]{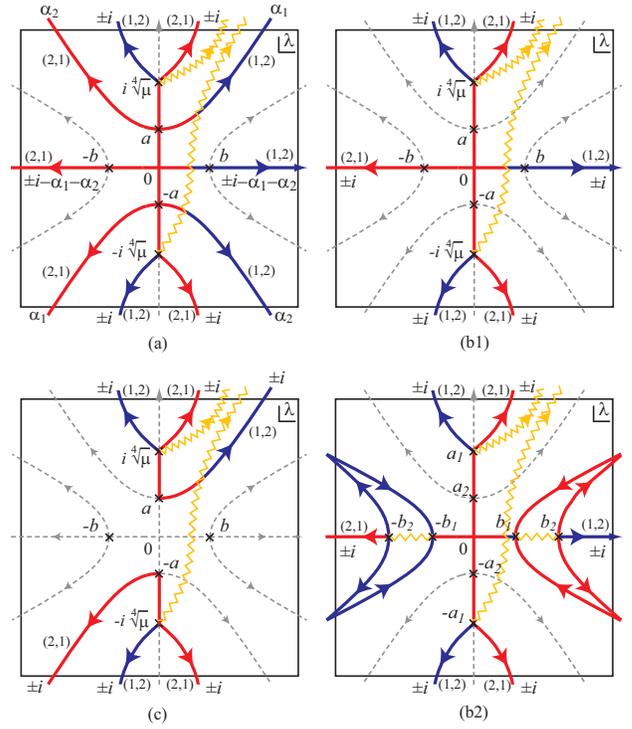}
\end{center}
\caption{\footnotesize The Deift-Zhou network for Yang-Lee edge. Here there are two saddle points in the string background $\varphi(\lambda)$: $a= i \sqrt{\sqrt{\mu}(\sqrt{5}-1)/4}$ and $b=\sqrt{\sqrt{\mu}(\sqrt{5}+1)/4}$. a) General solutions to the completion with two-cut boundary condition. b1) The solution with the single-line condition and hermiticity condition. c) Deformation of network around the essential singularity. This deformation changes the theory but helps us to obtain decay rates of perturbative string theory. b2) Deformation of spectral curve with keeping the same Stokes data as (b1). This gives the true vacuum if there is no large instanton along the network. }
\label{DZgraph25}
\end{figure}

\paragraph{$(p,q)=(2,5)$ minimal strings: Yang-Lee edge} \,\,\,
The DZ network of the spectral curve is drawn in Fig.~\ref{DZgraph25}-a. 
In this case, the hermiticity condition and the single-line condition are consistent with each others (Fig.~\ref{DZgraph25}-b1, as expected from matrix models). However, the RH 
integral picks up a exponentially-large-instanton contribution from the second saddle point 
($\lambda=\pm b$ in Fig.~\ref{DZgraph25}): $\mathcal F \asymeq \mathcal F_{\rm pert}(g;\mu) + \mathcal F_{\rm nonpert.}(g;\mu)$, 
\begin{align}
\mathcal F_{\rm nonpert.}(g;\mu) = \mp \frac{\sqrt{g/2}\exp\bigl[{+\frac{10}{21g} \sqrt{2(5-\sqrt{5})}\mu^{\frac{7}{4}}}\bigr]}{\sqrt{5\pi (2\sqrt{5})^{\frac{3}{2}} (\sqrt{5}+1)^{\frac{5}{2}}} \mu^{\frac{7}{8}}}+\cdots, 
\end{align}
with $u=g^2\del_t^2\mathcal F(t;\mu)$, $0=2t+2(g/2)^4u''''-5\mu u+5(gu'/2)^2+10 (g/2)^2u u'' + 5 u^3$ and $t \to 0$. 
This is the standard contribution from the $(1,2)$ {\em ghost} ZZ-brane (See also \cite{SatoTsuchiya,fisfim}).%
\footnote{For more about ZZ branes \cite{ZZ}, See also \cite{SeSh}. }
Note that $\mathcal F(g)$ is a real function and here is shown the leading of one-instanton contributions, and the multi-instanton contributions have stronger exponential behavior. By this, we conclude that the Yang-Lee edge perturbative string vacuum is unstable. More precisely, since there is no tachyon in its perturbative spectrum, this string theory vacuum is {\em meta-stable}. 

Generally meta-stable vacua in quantum systems have an important characteristic by {\em decay rate}.%
\footnote{Here we define {\em decay rates} by the imaginary part of ``energy'' of the meta-stable states in the following sense: $e^{\mathcal F} = \langle {\rm vac}|e^{-TH}|{\rm vac}\rangle \sim e^{-TE_{\rm vac}}\, (T\to \infty), E_{\rm vac} = E + i \Gamma_{\rm vac}$. Since minimal string theory is an Euclidean theory with ``compact Euclidian time'', our decay rate is simply given by imaginary part of the free-energy of meta-stable vacuum. Therefore, this definition/terminology can be easily generalized to the Lorenzian situations of string theory. }
Here, with use of the network, i.e.~path-integral degree of freedom in string theory, we caluculate the decay rate by applying the prescription of \cite{Coleman}. The way is to deform the path/network to avoid the instability (now by discarding hermiticity condition with keeping the single-line condition as in Fig.~\ref{DZgraph25}-c): 
\begin{align}
\mathcal F(g;\mu) \overset{\rm deform.}{\to} &\mathcal F^{\rm (def)}(g;\mu) \asymeq \mathcal F_{\rm pert}(g;\mu) + \mathcal F_{\rm nonpert.}^{\rm (def)}. 
\end{align}
Then its imaginary part is the {\em decay rate} of the Yang-Lee edge string vacuum ($g,\mu>0$): 
\begin{align}
{\rm Im}\,\mathcal F_{\rm nonpert.}^{\rm (def)} = \frac{\mp 1 }{2}\frac{\sqrt{g/2}\exp\bigl[{-\frac{10}{21g} \sqrt{2(5+\sqrt{5})}\mu^{\frac{7}{4}}}\bigr]}{\sqrt{5\pi (2\sqrt{5})^{\frac{3}{2}} (\sqrt{5}-1)^{\frac{5}{2}}} \mu^{\frac{7}{8}}}+\cdots. 
\end{align}
This is {\em one half of} the standard $(1,1)$ ZZ-brane contribution 
(See also \cite{SatoTsuchiya}). 

Generally one can see that the $(2,q)$ minimal string theory is meta-stable: 
The string theory with hermiticity has ghost $(1,2m)$ ZZ-branes ($m=1,2,\cdots,(q-3)/2$) which render the vacuum unstable; the decay rate is given by half of the $(1,1)$ ZZ-brane. String theory of pure-gravity, $(2,3)$, is an exception, since there is no ghost ZZ-brane in its background and therefore it is a stable vacuum. In this way, we have shown that physics of the landscape is given by the networks/fugacity which control the spectrum of (ghost) D-instantons. Importantly, this example shows that different networks (Fig.~\ref{DZgraph25}-b1 and \ref{DZgraph25}-c), i.e.~different fugacities, may represent different physical situations of the same theory.

\subsection{The true vacuum and landscape of string theory}
Since the Yang-Lee edge string theory is meta-stable, 
there is {\em the true vacuum}, into which the string theory decays. 
For that purpose, we choose the spectral curve in the landscape, 
\begin{align}
\varphi_{\rm tv}(\zeta) \in \mathfrak L_{\rm str} \qquad \bigl(\varphi_{\rm mstr}(\zeta) \in \mathfrak L_{\rm str}\bigr),
\end{align}
in such a way that there is large instantons along the Deift-Zhou network of the RH integral \eq{RHintegral}. Roughly speaking, the instanton actions on the saddle points $\lambda_*$ ($\del_\lambda\bigl[\varphi^{(j)}(\lambda^p_*)-\varphi^{(l)}(\lambda^p_*)\bigr] = 0$) should be positive real: 
\begin{align}
{\rm Re}\bigl[\varphi^{(j)}(\zeta_*)-\varphi^{(l)}(\zeta_*)\bigr]<0,
\end{align}
if the corresponding lines of the network $\mathcal K_m$ (with non-zero Stokes multiplier $s_{m,j,l}\neq 0$) are attached to the saddle point $\lambda^*$. In the current cases of $p=2$, it is eventually equivalent to vanishing condition around B-cycle 
on the network $\mathcal K$:
\begin{align}
\oint_{B} d\zeta\, \del_\zeta \bigl[\varphi_{\rm tv}^{(1)}(\zeta)- 
\varphi_{\rm tv}^{(2)}(\zeta)
\bigr]= 0\qquad B\subset \mathcal K, 
\end{align}
which is known as the Boutroux equations in the RH context \cite{ItsBook}. 
This kind of condition has been discussed also in old literatures of matrix models  \cite{David}. 
This simply means that the eigenvalues should fill up to the same Fermi-level of the effective potential along the DZ network. 
Here we simply show the result: 
\begin{align}
\zeta = \sqrt{\mu} \bigl(\wp (z)+c\bigr),\quad \del_\zeta \varphi^{(1)}_{\rm tv}(\zeta^{1/2}) = \sqrt{2 \mu^{\frac{5}{2}}}\bigl(\wp(z) - \alpha\bigr) \wp'(z). 
\end{align}
Here the Weierstrass $\wp$ function is given by $(\wp'(z))^2 = 4 (\wp(z))^3 -g_2 \wp(z)-g_3.$
The normalization of the system is now fixed as 
$\alpha = \frac{5}{2}c, \, g_2 = 5(1-3c^2),\, g_3 = 5c(2-7c^2)$ so that the corresponding string-background $\varphi_{\rm tr}(\zeta)$ belongs to the landscape $\mathfrak L_{\rm str}$. Therefore, the parameter $c$ is 
{\em an coordinate of the string theory landscape} of the Yang-Lee edge, 
and the true-vacuum condition is expressed with the Weierstrass elliptic functions: 
\begin{align}
 \Bigl[\frac{4 g_2}{5} \zeta_W(\omega_B)-\frac{6 g_3 \omega_B}{5}\Bigr] \alpha = \frac{6 g_3}{7}  \zeta_W(\omega_B)-\frac{g_2^2 \omega_B}{21},
\end{align}
where $\omega_B$ is the Weierstrass half period along the $B$-cycle, 
and $\zeta_W(z)$ is the Weierstrass $\zeta$ function, $\zeta_W'(z)=-\wp(z)$.
The numerical value of $c$ is given as $c \simeq -0.184963725\dots$. 
Then the perturbative amplitude around this true vacuum is obtained (by the RH approach with the network of Fig.~\ref{DZgraph25}-b2) as
\begin{align}
u(\mu) \simeq  - \sqrt{\mu}\Bigr(\wp(\omega_A)+\wp(\omega_B)-\wp(\omega_A+\omega_B) + c\Bigl).  
\end{align}
Here $\omega_A$ is the half period of the $A$-cycle. 

\begin{figure}[htbp]
\begin{center}
\includegraphics[scale=1]{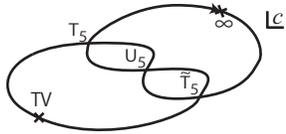}
\end{center}
\caption{\footnotesize One parameter landscape of Yang-Lee edge string theory. The coordinate is given by $c$. $T_5$, $\widetilde T_5$ and $U_5$ are backgrounds which are described by spectral curves with genus-zero. The value of $c$ for each vacuum is $c=\frac{1\pm \sqrt{5}}{6}, \pm \frac{1+\sqrt{5}}{6},  \frac{\pm \sqrt{5}}{3\sqrt{2}}$, respectively. }
\label{FigLandscape}
\end{figure}

The perturbative structure around the true vacuum does not receive any contributions from non-perturbative ambiguities, which is a result of universality. Note that, since the expression includes elliptic functions, this vacuum represents non-perturbative vacuum whose classical dynamics would not be stringy degree of freedom although quantum corrections still resembles the closed-string behavior $g^{2n-2}$. It would be worth drawing the string theory landscape with the parameter $c$ (Fig.~\ref{FigLandscape}).  The pinched points correspond to perturbative-string vacua ($T_5$, $\widetilde T_5$ and $U_5$) in which the perturbative amplitudes have simple the power scale behavior. $T_5$ is the original minimal-string vacuum \eq{SpectralCurveMinimalString}. Note that not all the vacua have a simple interpretation by matrix models and therefore most of the vacua are off-shell background of this non-perturbative string theory. 

These analyses can clearly be generalized to many other systems. Further investigations including general $(p,q)$ cases will be reported in future communication \cite{CIY5}.



\begin{acknowledgements}
The authors would like to thank M.~Fukuma, K.~Furuuchi, P.-M.~Ho, G.~Ishiki, A.~R.~Its, S.~Kawamoto, Yo.~Matsuo, T.~Nakatsu, R.~Schiappa, S.~Terashima and D.~Tomino for useful discussions and comments. The authors greatly thank P.-M.~Ho for kind supports on this collaboration which mostly takes place in NCTS north at NTU. 
C.-T.~Chan and H.~Irie are supported in part by National Science Council (NSC) of Taiwan under the contract 
No.~99-2112-M-029-001-MY3 (C.-T.~Chan) and No.~100-2119-M-007-001 (H.~Irie). 
H.~Irie is also supported in part by Japan Society for the Promotion of Science (JSPS). 
The authors are also supported in part by Taiwan String Theory Focus Group 
in NCTS under NSC No.~100-2119-M-002-001. 
\end{acknowledgements}



\end{document}